\documentstyle[floats,prd,aps,epsf,eqsecnum,12pt]{revtex}

\makeatletter
\newbox\tempboxa
\newdimen\captionboxsubcount
\def\capsize#1{\captionboxsubcount=#1pt}
\newdimen\captionboxsub
\captionboxsub=\hsize \advance\captionboxsub by -\captionboxsubcount
\advance\captionboxsub by -\captionboxsubcount
\long
\def\@makecaption#1#2{
\setbox\@tempboxa\hbox{#1 #2}
\ifdim \wd\@tempboxa >\captionboxsub
\rightskip=\captionboxsubcount \leftskip=\captionboxsubcount #1 #2
\else \hbox to\hsize{\hfil\box\@tempboxa\hfil}
\fi}
\makeatother
\capsize{30}

\begin{document}

\begin{titlepage}
\begin{flushright}
\begin{minipage}{5cm}
\begin{flushleft}
\small
\baselineskip = 13pt
YCTP-P7-99\\ TMUP-HEL-9905\\ hep-th/9905521 \\
\end{flushleft}
\end{minipage}
\end{flushright}
\begin{center}
\Large\bf
Competing Condensates in Two Dimensions
\end{center}
\vfil

\begin{center}

Alan {\sc Chodos}\footnote{ Electronic address : {\tt
chodos@hepvms.physics.yale.edu}}\\ { \it \qquad Department of
Physics, Yale University, New Haven, CT 06520-8120}\\

Fred {\sc Cooper}\footnote{Electronic address: {\tt
cooper@schwinger.lanl.gov}}
\\ {\it \qquad Los Alamos National Laboratory, Los Alamos, NM 87545} \\

Hisakazu {\sc Minakata}\footnote{Electronic address: {\tt
minakata@phys.metro-u.ac.jp}} \\
{\it \qquad Department of Physics, Tokyo Metropolitan University\\
Minami-Osawa, Hachioji, Tokyo 192-0397, Japan and\\
Research Center for Cosmic Neutrinos, Institute for Cosmic Ray Research\\
University of Tokyo, Tanashi, Tokyo 188-8502, Japan}

\vskip .5cm
 {\sc }

\qquad

\end{center}
\vfill
\begin{center}
\bf
Abstract
\end{center}
\begin{abstract} We generalize our previous $2$-dimensional model in which
a pairing condensate
 $\langle \psi \psi \rangle$ was generated at large $N$.
In the present case, we allow for both $\langle \psi\psi
\rangle$ and a chiral condensate $\langle \bar{\psi}\psi \rangle$ to exist.
 We construct the effective potential to leading order in $1 \over N$, and
derive the gap equations at finite density and temperature. We
study the zero density and temperature situation analytically. We
perform the renormalization explicitly and we show that the physics
is controlled by a parameter related to the relative strengths of
the interactions in the pairing and chiral channels. We show that
although a solution to the gap equations exists in which both
condensates are non-vanishing, the global minimum of the effective
potential always occurs for the case when one or the other
condensate vanishes.

\baselineskip = 17pt

\end{abstract}
\begin{flushleft}

PACS numbers:
\end{flushleft}
\vfill
\end{titlepage}

\section{Introduction}

\bigskip
In a previous paper [1], we introduced a variant of the Gross-Neveu
model which, in the large $N$ limit, exhibits the formation of a
pairing condensate $\langle \psi \psi \rangle$. In that work, we
derived and solved the gap equation, demonstrated that the coupling
was asymptotically free, and discussed some of the properties of
the relevant Green's functions. Our work was motivated, at least in
part, by recent interest in the formation of similar condensates in
$QCD$ in the presence of a chemical potential. [2-4]

But in real $QCD$, as well as in a variety of condensed-matter
systems, the pairing condensate is not the only possible one. In
$QCD$ at low density and temperature there is a chiral condensate;
it is only as the density increases that one believes the $\langle
\bar{q}q \rangle$ condensate disappears and a new phase, characterized by a
non-vanishing $\langle qq \rangle$, replaces it. Likewise, in other
systems, which condensates exist depends on external parameters
like the density and temperature, and also on the relative
strengths of various couplings. With this in mind, we extend our
previous work by considering a more general model governed by two
independent couplings: ~the original Gross-Neveu term [5] that
promotes the condensation of $\langle \bar{\psi}\psi \rangle$, and
the term considered in our earlier paper that produces a $\langle
\psi \psi \rangle$ condensate. As in ref. $1$, we shall be able to write
down the gap equations exactly to leading order
in $1 \over N$, and to renormalize the couplings, thereby rendering
the gap equations finite. For the general case of non-vanishing
density and temperature, we shall be able to write the gap
equations in terms of a single integral over a momentum variable
$k$; the effective potential is then expressible as a further
integral over the auxiliary fields. To solve the gap equations will
require numerical evaluation of these integrals, which takes us
beyond the scope of the present work.

If we set temperature and chemical potential to zero, we can derive
a closed-form expression for the effective potential, and can
explicitly perform the renormalization of the couplings. We then
find that there is one dimensionless parameter, independent of the
renormalization scale, whose value determines which of the
condensates is present. This situation might be described as
"partial dimensional transmutation": ~the unrenormalized theory has
two bare couplings whereas the renormalized one has a
renormalization scale, which is arbitrary, and a dimensionless
parameter, independent of this scale that controls the physics. We
find that the gap equations have three types of solution: ~two in
which one or the other of the condensates vanish, and a third,
mixed case, in which both condensates are non-vanishing. It turns
out, however, that the true minimum of the effective potential is
always at a point where one of the condensates vanishes. Thus the
mixed case is never realized physically, at least for zero
temperature and density.

In the remainder of the paper, we shall describe the analysis that
leads to the above results. In section II, we define the model and
derive the gap equations for the general case of non-vanishing
density and temperature. In section III, we specialize to the case
$\mu = T = 0$, and derive an explicit form for the effective
potential. We renormalize the coupling constants and derive thereby
the renormalized effective potential and gap equations. Section IV
is devoted to an analysis of the gap equation and we derive the
conditions under which one or the other condensate dominates.
Section V contains some conclusions.

\section{General Considerations}

\bigskip
We begin with the Lagrangian:

\begin{eqnarray}
{\cal L} &=& \bar{\psi}^{(i)} i \bigtriangledown  \!\!\!\!\!\! /
\psi^{(i)} + {1
\over 2} g^2 [\bar{\psi}^{(i)} \psi^{(i)}][\bar{\psi}^{(j)} \psi^{(j)}]
\nonumber \\
&+& 2 G^2 (\bar{\psi}^{(i)} \gamma_5 \psi^{(j)})(\bar{\psi}^{(i)}
\gamma_5 \psi^{(j)}) - \mu \psi^{\dagger (i)} \psi^{(i)} .
\end{eqnarray}

\noindent
The flavor indices, summed on from $1$ to $N$, have been explicitly
indicated. The first quartic term is the usual Gross-Neveu
interaction, whereas the second such term, which differs in the
arrangement of its flavor indices, induces the pairing force to
leading order in ${1 \over N}$. In the final term, $\mu$ is the
chemical potential.

Strictly speaking, a $\langle \psi \psi \rangle$ condensate cannot
form, because it breaks the $U(1)$ of Fermion number and hence
violates Coleman's theorem [6]. Similarly,
$\langle \bar{\psi} \psi \rangle$ as well as $\langle \psi \psi \rangle$
condensates cannot exist at finite temperature in one spatial dimension
because of the Mermin-Wagner theorem [6].  Nevertheless, it is
meaningful to study the formation of such condensates to leading
order in ${1 \over N}$, as explained in ref. [7].
%

Our conventions are: ~$\gamma^0 = \sigma_1$; $\gamma^1 = - i
\sigma_2$; $\gamma_5 = \sigma_3$. The pairing term, proportional to $G^2$,
may then be rewritten:

\begin{eqnarray}
2G^2 \bar{\psi}^{(i)} \gamma_5 \psi^{(j)} \bar{\psi}^{(i)} \gamma_5
\psi^{(j)} = - G^2 [\epsilon_{\alpha\beta} \psi_{\alpha}^{\dagger (i)}
\psi_{\beta}^{\dagger (i)}]
[\epsilon_{\gamma \delta} \psi_{\gamma}^{(j)}
\psi_{\delta}^{(j)}] ~.
\end{eqnarray}

\bigskip\noindent
Following standard techniques [8] we add the following terms
involving auxiliary fields $m$, $B^{\dagger}$, and $B$:

\begin{eqnarray}
\bigtriangleup {\cal L} = - {1 \over 2 g^2} [m + g^2 \bar{\psi} \psi]^2 -
{1 \over G^2} (B^{\dagger} - G^2 \epsilon_{\alpha\beta}
\psi_{\alpha}^{\dagger (i)} \psi_{\beta}^{\dagger (i)})(B + G^2
\epsilon_{\gamma \delta} \psi_{\gamma}^{(j)} \psi_{\delta}^{(j)}) ~.
\end{eqnarray}

\noindent
This addition to ${\cal L}$ will not affect the dynamics. In ${\cal
L}^{\prime}
= {\cal L} + \bigtriangleup {\cal L}$, the terms quartic in fermion
fields cancel, and we have

\begin{eqnarray}
{\cal L}^{\prime} = \bar{\psi} (i \bigtriangledown  \!\!\!\!\!\! /
- m - \mu
\gamma^0) \psi - {m^2 \over 2 g^2} - {B^{\dagger}B \over G^2} + B
\epsilon_{\alpha\beta} \psi_{\alpha}^{\dagger (i)}
\psi_{\beta}^{\dagger (i)} - B^{\dagger} \epsilon_{\alpha\beta}
\psi_{\alpha}^{(i)} \psi_{\beta}^{(i)} ~.
\end{eqnarray}

\bigskip\noindent
We integrate out $\psi$ and $\psi^{\dagger}$ to obtain the
effective action depending on the auxiliary fields $m$, $B$ and
$B^{\dagger}$:

\begin{eqnarray}
\Gamma_{eff} (m, B, B^{\dagger}) = \int d^4x (- {m^2 \over 2 g^2} -
{B^{\dagger}B \over G^2}) - {i \over 2} Tr \log A^T A - {i \over 2} Tr \log
[{\bf 1} + M^2 (A^T)^{-1} \sigma_2
A^{-1} \sigma_2]
\end{eqnarray}

\bigskip\noindent
where we have subtracted a constant (independent of the auxiliary
fields) and have defined:

\begin{eqnarray}
A = \gamma^0 (i \bigtriangledown  \!\!\!\!\!\! / - m - \mu
\gamma^0) = i \partial_0 + i \sigma_3 \partial_x - \mu - m \sigma_1
\end{eqnarray}

\bigskip\noindent
so that $A^T = - i \partial_0 - i \sigma_3 \partial_x - \mu - m
\sigma_1$.

Since we are looking for a vacuum solution, we have assumed in
(2.5) that $B, B^{\dagger}$ and $m$ are constants and have set $M^2
= 4 B^{\dagger}B$. The trace on flavor indices will give a factor
$N$. The large-$N$ limit is achieved by setting $g^2N
= \lambda$, and $G^2N = \kappa/4$, and letting $N \rightarrow \infty$ with
$\lambda$ and $\kappa$ fixed.
We define the effective potential $V_{eff}$ via

\begin{eqnarray}
\Gamma_{eff} = - N(\int d^2x) V_{eff}
\end{eqnarray}

\bigskip\noindent
and we therefore have

\begin{eqnarray}
V_{eff} (m, M) = {m^2 \over 2 \lambda} + {M^2 \over \kappa} +
V_{eff}^{(1)} (m, M) ~,
\end{eqnarray}

\bigskip\noindent
with $V_{eff}^{(1)} (m, M) = {i \over 2} [tr \log (A^TA)_{xx} + tr
\log ({\bf 1} + M^2(A^T)^{-1} \sigma_2 A^{-1} \sigma_2)_{xx}]$, where
now the trace is only over the spinor indices.

We next generate the local extrema of $V_{eff}$ by solving

\begin{eqnarray}
{\partial V_{eff} \over \partial m^2} = {\partial V_{eff} \over
\partial M^2} = 0 ~.
\end{eqnarray}

\bigskip\noindent
We evaluate the matrix products in $V_{eff}^{(1)}$ in momentum
space, with $\partial_{\mu} \rightarrow i k_{\mu}$. The traces can
be done with the help of

\begin{eqnarray}
tr [{1 \over V_0 + \vec{V} \cdot \vec{\sigma}}] = {2 V_0 \over
V_0^2 - \vec{V}^2}
\end{eqnarray}

\bigskip\noindent
for any $V_0, \vec{V}$. After some manipulation, equations (2.9)
become

\begin{eqnarray}
{1 \over 2 \lambda} = - {\partial V_{eff}^{(1)} \over \partial m^2}
= i \int {d^2 k \over (2 \pi)^2} {[k_0^2 - k_1^2 + \mu^2 + M^2 - m^2] \over D}
\end{eqnarray}

\begin{eqnarray}
{1 \over \kappa} = - {\partial V_{eff}^{(1)} \over \partial M^2}
= i \int {d^2 k \over (2 \pi)^2} {[k_0^2 - k_1^2 - \mu^2 - M^2 + m^2] \over D}
\end{eqnarray}

\bigskip\noindent
where $D = [k_0^2 - k_1^2 - M^2 + m^2 - \mu^2]^2 - 4[m^2 k_0^2 + \mu^2
k_1^2 - m^2 k_1^2]$. In this expression, $k_0$ is shorthand for
$k_0 + i\epsilon sgn k_0$, where $\epsilon \rightarrow 0^+$. This
prescription correctly implements the role of $\mu$ as the chemical
potential.

The equations can be reduced further by doing the $k_0$ integral.
Let us define $k_{\pm} = \sqrt{b_1 \pm 2 b_2}$, where $b_1 = M^2  +
m^2 + \mu^2 + k_1^2$, and $b_2 = [M^2 m^2 + \mu^2(k_1^2 + m^2)]^{{1
\over 2}}$. Then evaluating the $k_0$ integral by contour methods,
taking proper account of the $i\epsilon$ prescription mentioned
above, we find

\begin{eqnarray}
{1 \over 2\lambda} = {1 \over 8\pi} \int_{- \Lambda}^{\Lambda} dk_1
[{1 \over k_+} + {1 \over k_-} + {(M^2 + \mu^2) \over \sqrt{M^2  m^2 +
\mu^2 (k_1^2 + m^2)}} ({1 \over k_+} - {1 \over k_-})]
\end{eqnarray}

\noindent
and

\begin{eqnarray}
{1 \over \kappa} = {1 \over 8\pi} \int_{- \Lambda}^{\Lambda} dk_1
[{1
\over k_+} + {1 \over k_-} + {m^2 \over \sqrt{M^2 m^2 +
\mu^2 (k_1^2 + m^2)}} ({1 \over k_+} - {1 \over k_-})] ~.
\end{eqnarray}

\bigskip\noindent
The $k_1$ integrals are logarithmically divergent and we have
regularized them by imposing a cutoff $\Lambda$. This will be
absorbed in the renormalization process to be described in the next
section. Note, however, that the combination ${1 \over 2\lambda} -
{1 \over \kappa}$ is given by a convergent integral. This fact will
ultimately lead to the renormalization-scale independent constant
mentioned in the introduction.

We observe from the form of equations (2.11) and (2.12) that the
function $V_{eff}^{(1)}$ can be reconstructed by integrating with
respect to $m^2$ and $M^2$ in the expressions for ${1 \over 2
\lambda}$ and ${1
\over \kappa}$. This will determine $V_{eff}^{(1)}$ up to a single constant
$V_{eff}^{(1)}(0,0)$, which can be chosen arbitrarily without
affecting any physical quantity. Explicitly performing this
integration we obtain for the unrenormalized
 determinant correction to the
effective potential

\begin{equation}
V^{(1)} (m,M) = -{1 \over 2 \pi} \int_0^{\Lambda} dk_1 [ k_+ + k_{-} ]
\end{equation}

\bigskip

To generalize this discussion to the case of non-zero temperature,
one returns to eqns. (2.11) and (2.12), and one continues to
Euclidean space via the replacement $k_0
\rightarrow  -i k_4$ with $k_4$ now considered real.
The statistical-mechanical partition function is obtained from the
Euclidean zero temperature path integral by integrating over a
finite regime in imaginary time $\tau = it$ from $0$ to $\beta = {1
\over kT}$. Because of the cyclic property of the trace, the
Fermion Green's functions are anti-periodic in $\tau$ and one has
the replacement

\begin{equation}
\int dk_4 \rightarrow  {2 \pi \over \beta} \sum_{n}
\end{equation}

\noindent
where the antiperiodicity gives the Matsubara frequencies:

\begin{equation}
\omega_n = k_{4_n} = {(2 n+1) \pi \over \beta}
\end{equation}

To do the sum over the Matsubara frequencies, one uses the calculus of
residues to obtain the identity:

\begin{equation}
 {2 \over \beta} \sum_n f(i \omega_n) = - \sum_s \tanh {\beta z_s \over 2}~~
 Res f(z_s)
\end{equation}

\noindent
where $z_s$ are the poles of $f(z)$ in $z$ in the complex plane; $
Res f(z_s)$ is the residue of $f(z)$ at $z_s$ and we have assumed
the function $f(z)$ falls off at least as fast as
$1/|z|^{1+\epsilon}$ for large $\mid z \mid$. It will be convenient
to use:

\[   \tanh {\beta z_s \over 2} = 1- 2 n_f(z_s) \]

\noindent
where
\[
   n_f(z) = {1 \over e^{\beta z} +1 } \]

\noindent
is the usual Fermi-Dirac distribution function.

Rotating equations (2.11) and (2.12) into Euclidean space as
described above, we get:

$$
{1 \over 2 \lambda} = - {\partial V_{eff}^{(1)} \over \partial m^2}
=  \int {d^2 k \over (2 \pi)^2} {[-k_4^2 - k_1^2 + \mu^2 + M^2 - m^2] \over D}
$$

$$
{1 \over \kappa} = - {\partial V_{eff}^{(1)} \over \partial M^2}
=  \int {d^2 k \over (2 \pi)^2} {[-k_4^2 - k_1^2 - \mu^2 - M^2 + m^2] \over D}
$$

\noindent
where now

\begin{equation}
\int dk_4 \equiv   {2 \pi \over \beta} \sum_{n}
\end{equation}

\bigskip\noindent
where $D = [-k_4^2 - k_1^2 - M^2 + m^2 - \mu^2]^2 - 4[- m^2 k_4^2 +
\mu k_1^2 - m^2 k_1^2]$. There is no longer any need for an $i \epsilon $
in the definition of $k_4$. Performing the sums over the Matsubara
frequencies we obtain the unrenormalized form of the equations
which are given by the same expression as the zero temperature ones
found earlier, with the replacements:

\begin{eqnarray}
  {1 \over k_{+}} \rightarrow && {1 \over k_{+}} ( 1- 2 n_f (k_{+}))
\nonumber \\
  {1 \over k_{-}} \rightarrow && {1 \over k_{-}} ( 1- 2 n_f (k_{-}))
\end{eqnarray}

As before we can integrate this to get the determinant correction to the
effective potential which in unrenormalized form is:

\begin{equation}
V^{(1)} (m,M) = -{1 \over 2 \pi} \int_0^{\Lambda} dk_1 [ k_+ + k_- + {2
\over \beta} \log (1 + e^{-\beta k_+}) + {2 \over \beta} \log (1 +
e^{-\beta k_-}) ]
\end{equation}

\section{The case $\mu = T = 0$}

\bigskip
Renormalization of the effective potential is best discussed in the
context of the zero temperature and density sector of the theory
where we can define the renormalized coupling constant in terms of
the physical scattering of Fermions at a particular momentum scale.
This vacuum sector is interesting in its own right and we shall be
able, by analytic means, to derive the result that depending on the
relative strengths of the 2 couplings the theory will be in one or
another broken phase but never in a mixed phase. Setting $\mu = T=
0$ we obtain

\begin{equation}
{\partial V_{eff}^{(1)} \over \partial m^2} = - {1 \over 4 \pi}
\int_0^{\Lambda}dk_1 [(1 + {M \over m}) {1 \over \sqrt{k_1^2
 + (m + M)^2}} + (1 - {M \over m}) {1 \over \sqrt{k_1^2 + (m
- M})^2]}
\end{equation}

\begin{equation}
{\partial V_{eff}^{(1)} \over \partial M^2} = - {1 \over 4 \pi}
\int_0^{\Lambda}dk_1 [(1 + {m \over M}) {1 \over \sqrt{k_1^2
 + (m + M)^2}} + (1 - {m \over M}) {1 \over \sqrt{k_1^2 + (m
- M})^2]}
\end{equation}

\bigskip\noindent
which is solved by

\begin{equation}
V^{(1)}(m, M) = - {1 \over 2\pi} \int_0^{\Lambda}dk_1 [\sqrt{k_1^2
 + (M + m)^2} + \sqrt{k_1^2
 + (M - m)^2} - 2 k_1] ~.
\end{equation}

\bigskip\noindent
This can be integrated to give the unrenormalized effective
potential:

\begin{eqnarray}
V_{eff}(m, M) &=& M^2 [{1 \over \kappa} - {1 \over 4\pi}] + m^2 [{1
\over 2\lambda} - {1 \over 4\pi}] \nonumber \\
&-& {1 \over 4 \pi} [(M + m)^2 ln ({2 \Lambda \over M + m}) + (M -
m)^2 ln ({2 \Lambda \over \mid M - m
\mid})] ~.
\end{eqnarray}

\bigskip\noindent
We renormalize by demanding that the renormalized couplings
$\kappa_R$ and $\lambda_R$ satisfy

\begin{equation}
{\partial^2 V_{eff} \over \partial B \partial
B^{\dagger}}\mid_{\stackrel{M
= M_0}{m = m_0}} = {4 \over \kappa_R}
\end{equation}

\noindent
and

\begin{equation}
{\partial^2 V_{eff} \over \partial m^2}\mid_{\stackrel{M
= M_0}{m = m_0}} = {1 \over \lambda_R} ~.
\end{equation}

\bigskip\noindent
Here $M = M_0$, $m = m_0$ designates an arbitrary renormalization
point on which the couplings will depend. Using these conditions to
solve for $\lambda$ and $\kappa$ in terms of $\lambda_R$ and
$\kappa_R$ yields the renormalized form of the effective potential:

\begin{eqnarray}
V_{eff} = m^2 [a &+& {1 \over 4 \pi} ln \mid {M^2 - m^2 \over
\gamma_0}
\mid ] + M^2 [b + {1 \over 4 \pi} ln \mid {M^2 - m^2 \over \gamma_0} \mid ]
\nonumber \\
&+& {1 \over 2 \pi} m M ln \mid {M + m \over M
- m} \mid
\end{eqnarray}

\bigskip\noindent
where $a$ and $b$ are the following constants:

\begin{eqnarray}
a &=& {1 \over 2 \lambda_R} - {3 \over 4 \pi} \nonumber \\ &~~&
\nonumber
\\ b &=& {1
\over
\kappa_R} - {1 \over 2 \pi} + {1 \over 8 \pi} {m_0 \over M_0} ln
\mid {M_0 - m_0 \over M_0 + m_0} \mid
\end{eqnarray}

\noindent
and $\gamma_0 = \mid M_0^2 - m_0^2 \mid$.

Note that the renormalization we have just performed at $\mu = T =
0$ is also sufficient to remove all divergences from the effective
potential in the more general case of non-vanishing chemical
potential and temperature. The addition of $\mu$ and $T$ will only
result in finite corrections to the gap equations and therefore to
the vacuum values of $m$ and $M$. We shall return to this point in
section V.

The gap equations are properly derived by differentiating $V_{eff}$
with respect to $B$ and $m$ and then setting these derivatives to
zero. Because $V_{eff}$ depends only on $B^{\dagger}B$ and $m^2$,
it will always be possible to have solutions with one of $m$ or $B$
or perhaps both set to zero. The gap equations can be written

\begin{equation}
m [2a + {1 \over 2 \pi} + {1 \over 2 \pi} ln {\mid M^2 - m^2 \mid
\over \gamma_0}] - {M \over 2 \pi} ~ln \mid {M - m \over
M + m} \mid = 0
\end{equation}

\noindent
and

\begin{equation}
M [b + {1 \over 4 \pi} + {1 \over 4 \pi} ln {\mid M^2 - m^2 \mid
\over \gamma_0}] - {m \over 4 \pi} ln \mid {M - m \over M +
m} \mid = 0 ~.
\end{equation}

\bigskip\noindent
The solutions $m = m^*$ and $M = M^*$ will give us the local
extrema of $V_{eff}$. The first of these equations is an identity
if $m
= 0$, and the second if $M = 0$. Also, the values of $m$ and $M$
that solve these equations are physical parameters that must be
independent of the renormalization scale $\gamma_0$. Thus these
equations tell us how $a$ and $b$ individually run with $\gamma_0$.
We note, however, that if we solve for the combination

\begin{equation}
\delta = a - b = {1 \over 4 \pi} [{M^* \over m^*} - {m^* \over M^*}] ln
\mid {M^* - m^* \over M^* + m^*} \mid
\end{equation}

\bigskip\noindent
the scale $\gamma_0$ drops out. Therefore $\delta$ is a true
physical parameter in the theory; we shall see in the next section
that its value controls which of the two condensates $m$ and $M$
can exist.

\bigskip
\section{Analysis of the gap equations}

It will be useful in the following to note that, at a solution of
the gap equations (3.9) and (3.10), the effective potential takes
the simple form

\begin{equation}
V_{eff} (m, M) = - {1 \over 4 \pi} (m^2 + M^2) ~.
\end{equation}

\bigskip\noindent
Our goal is to analyze all the solutions of the gap equations and
to find the one that produces the global minimum of $V_{eff}$. This
will then represent the true vacuum of the theory.

There are four types of solution to (3.9) and (3.10). The first is
simply to set $m = M = 0$, leading of course to $V = 0$. Clearly,
from (4.1) we see that if any other solution exists, $V = 0$ cannot
be the minimum of $V$. The second and third types are obtained by
setting $M = 0$, $m \neq 0$ and $m = 0$, $M \neq 0$ respectively.
If $M = 0$, then from (3.9), we have

\begin{equation}
m^2 = \gamma_0 ~exp[ - (1 + 4 \pi a)]
\end{equation}

\noindent
so

\begin{equation}
V_0 (m, M = 0)= - {\gamma_0 \over 4 \pi} e^{-(1 + 4 \pi a)}
\end{equation}

\bigskip\noindent
(we shall use $V_0$ to denote values of $V_{eff}$ at solutions of
the gap eqn.). Likewise, if $m = 0, M \neq 0$, then from (3.10)

\begin{equation}
M^2 = \gamma_0 ~exp [ - (1 + 4 \pi b)]
\end{equation}

\begin{equation}
V_0 (m = 0, M) = - {\gamma_0 \over 4 \pi} ~e^{- (1 + 4 \pi b)} ~.
\end{equation}

\bigskip\noindent
Thus we see that

\begin{equation}
V_0 (m = 0, M) < V_0 (m, M = 0) ~~~{\rm if} ~\delta > 0
\end{equation}

\noindent
and

\begin{equation}
V_0 (m, M = 0) < V_0 (m = 0, M) ~~~{\rm if} ~\delta < 0 ~.
\end{equation}

The fourth case is when both $m$ and $M$ are non-vanishing. It is
then convenient to define $\rho = {M \over m}$ and to combine the
gap equations in the form

\begin{equation}
\delta = (\rho^2 - 1) [b + {1 \over 4 \pi} + {1 \over 4 \pi} ln ({m^2
\mid {\rho^2 - 1}\mid \over \gamma_0})]
\end{equation}

\noindent
and

\begin{equation}
\delta = {1 \over 4 \pi} {(\rho^2 - 1) \over \rho} ~ln~ \mid {\rho - 1
\over \rho + 1} \mid ~.
\end{equation}

\bigskip\noindent
Both these equations are even in $\rho$, so we may take $\rho > 0$
for convenience. Eqn. (4.9) tells us immediately that if $\delta >
0$, $0 < \rho < 1$, and if $\delta < 0$, $\rho > 1$. Furthermore,
the r.h.s. of (4.9) is bounded between $- {1 \over 2 \pi}$ and ${1
\over 2 \pi}$. Hence we conclude: ~If $\mid \delta \mid > {1 \over 2 \pi}$
 there is no solution with both $m$ and $M$ non-vanishing. If $\mid \delta
\mid < {1 \over 2 \pi}$,
 there is such a solution, with the property that $m > M$ if $\delta > 0$
and $M < m$ if $\delta < 0$.

It remains to decide whether $V_0(m, M)$ can be the global minimum.
To this end, it is convenient to reexpress the gap equations once
more in the following form:

\begin{equation}
- (1 + 4 \pi a) = ln ~{m^2 \mid \rho^2 - 1 \mid \over \gamma_0} - ln \{
~\mid {\rho - 1 \over \rho + 1} \mid^{\rho} \}
\end{equation}

\noindent
and

\begin{equation}
- (1 + 4 \pi b) = ln ~{m^2 \mid \rho^2 - 1 \mid \over \gamma_0} - ln \{
~\mid {\rho - 1 \over \rho + 1} \mid^{{1 \over \rho}} \}~.
\end{equation}

\bigskip\noindent
From these, making use of eqs. (4.1), (4.3) and (4.5), we
immediately obtain

\begin{equation}
V_0 (m = 0, M) = - {\gamma_0 \over 4 \pi} ~e^{- (1 + 4 \pi b)} =
g_1(\rho) V_0(m, M)
\end{equation}

\noindent
and

\begin{equation}
V_0 (m, M = 0) = - {\gamma_0 \over 4 \pi} ~e^{- (1 + 4 \pi a)} =
g_2(\rho) V_0(m, M)
\end{equation}

\noindent
where

\begin{equation}
g_1(\rho) = {(1 + \rho)^{1 + {1 \over \rho}} \mid 1 - \rho
\mid^{1
- {1 \over \rho}} \over 1 + \rho^2}
\end{equation}

\noindent
and

\begin{equation}
g_2(\rho) = {(\rho + 1)^{\rho + 1} \mid \rho - 1
\mid^{1
-  \rho} \over 1 + \rho^2} ~.
\end{equation}

\bigskip\noindent
Eq. (4.12) is the relevant comparison if ${1 \over 2 \pi} > \delta
> 0$ and $0 < \rho < 1$, whereas eq. (4.13) is relevant for $0 >
\delta
> - {1 \over 2 \pi}$ and $\rho > 1$.

We observe, however, that $g_2({1 \over \rho}) = g_1(\rho)$, so
both cases reduce to the following: ~if we can show that $g_1(\rho)
> 1$ in the range $0 < \rho < 1$, then $V_0(m, M)$ is never the global minimum
(recall that the $V_0's$ are all $< 0$). On the other hand, if
$g_1(\rho) < 1$ in this range, it will be possible to have $V_0(m,
M)$ be the global minimum.

To settle this question, write $g_1 = e^h$, with

\begin{eqnarray}
h(\rho) &=& (1 + {1 \over \rho}) ~ln~ (1 + \rho) + (1 - {1 \over
\rho}) ~ln~ (1 - \rho) - ~ln~ (1 + \rho^2) \nonumber \\
&=& ln~ [{1 + \rho \over 1 + \rho^2}] + {1 \over \rho} ~ln~ (1 +
\rho) + (1 - {1 \over \rho}) ~ln~ (1 - \rho) ~.
\end{eqnarray}

\bigskip\noindent
In the range of interest, $\rho^2 < \rho$, so the r.h.s. is a sum
of positive terms. Hence $h(\rho) > 0$ and $g_1(\rho) > 1$.

We conclude that the global minimum of $V_{eff}$ has $M = 0, m \neq
0$ if $\delta < 0$, and $m = 0, M \neq 0$ if $\delta > 0$.

\section{Conclusions}

From eqn. (2.21) we can see that the corrections due to
non-vanishing temperature and density do not affect the ultraviolet
behavior of the integrand in the $k_1$ integral defining $V^{(1)}$.
Therefore, the renormalization that we have performed at $\mu = T =
0$ in section III suffices to remove the ultraviolet divergences
from the effective potential, and will allow us to send the cutoff
to infinity. It is perhaps worth recording the complete result
explicitly. We find, from eqns. (3.5) and (3.6), that

\begin{equation}
{1 \over 2 \lambda} = a + {1 \over 4 \pi} + X
\end{equation}

\begin{equation}
{1 \over \kappa} = b + {1 \over 4 \pi} + X
\end{equation}

\bigskip\noindent
where $a$ and $b$ are defined by eqn. (3.8), and $X$ is a divergent
integral given by

\begin{eqnarray}
X &=& {1 \over 4 \pi} \int_0^{\Lambda} dk_1 [{1 \over \sqrt{k_1^2 +
(m_0 + M_0)^2}} + {1 \over \sqrt{k_1^2 + (m_0 - M_0)^2}}]
\nonumber \\
&=& {1 \over 2 \pi} ~[log ~({2 \Lambda \over \sqrt{\gamma_0}})]+
~{\rm terms ~which ~vanish ~as}~ \Lambda \rightarrow \infty ~.
\end{eqnarray}

\noindent
Thus the full renormalized effective potential may be written

\begin{eqnarray}
V_{eff} &=& \alpha_1 m^2 + \alpha_2 M^2 - {1 \over 2 \pi}
\int_0^{\infty} dk_1 [k_+ + k_- + {2 \over \beta} log~ (1 + e^{-\beta k_+})
\nonumber \\
&+& {2 \over \beta} log~ (1 + e^{- \beta k_-}) \nonumber \\ &-&
2k_1
-({m^2 + M^2 \over 2}) ({1 \over \sqrt{k_1^2 + (m_0 + M_0)^2}} + {1
\over \sqrt{k_1^2 + (m_0 - M_0)^2}})] ~.
\end{eqnarray}

\bigskip\noindent
where $\alpha_1 = {1 \over 4 \pi} (1 + 4 \pi a)$ and $\alpha_2 = {1
\over 4 \pi} (1 + 4 \pi b)$. If $\alpha_1 < \alpha_2$, then at $\mu = T =
0$ the vacuum has
$m^2 = m_F^2 \equiv \gamma_0 e^{-4 \pi \alpha_1} ~~{\rm and} ~~M^2
= 0.$ Here $m_F$ is the dynamically generated fermion mass. It is
convenient to choose the renormalization scale so that $m_F^2 =
\gamma_0$. This entails $\alpha_1 = 0, \alpha_2 > 0$. Furthermore,
 we are free to choose $M_0 = 0$, so that $m_0 = m_F$. Then $V_{eff}$ takes
the form

\begin{eqnarray}
V_{eff} = \alpha_2 M^2 &-& {1 \over 2 \pi} \int_0^{\infty} dk_1
[k_+ + k_- + {2 \over \beta} (log ~(1 + e^{- \beta k_+}) + log ~(1
+ e^{- \beta k_-})) - 2 k_1 \nonumber \\ &-& (m^2 + M^2) {1 \over
\sqrt{k_1^2 + m_F^2}}] ~.
\end{eqnarray}

\bigskip
We observe that if we set $M = 0$ in this expression, we obtain,
with $E = \sqrt{k^2 + m^2}$,

\begin{eqnarray}
V_{eff} (m^2, T, \mu) &=& {m^2 \over 4 \pi} [log~ {m^2 \over m_F^2}
- 1] \nonumber \\ &~& \nonumber \\ &-& {2 \over \beta}
\int_0^{\infty} {dk
\over 2
\pi} ~log~ [(1 + e
^{- \beta (E + \mu)})(1 + e^{- \beta (E - \mu)})]
\end{eqnarray}

\bigskip\noindent
which is the effective potential for the Gross-Neveu model in
agreement with refs. [9]  and [10], and furthermore at $T = 0$ the
integral can be done to give an explicit form that agrees with the
results of ref. [11].

\bigskip Similarly, in the opposite case $\alpha_2 <
\alpha_1$, we have, in the $\mu = T = 0$ vacuum, $m^2 = 0$ and $M^2
= \Delta^2
\equiv
\gamma_0 e^{- 4 \pi \alpha_2}$, where $\Delta$ is the dynamically generated
gap. So we
choose  $\alpha_2 = 0, \alpha_1 > 0$, and $m_0 = 0$, $\Delta^2 =
\gamma_0 = M_0^2$. The effective potential becomes

\begin{eqnarray}
V_{eff} = \alpha_1 m^2 &-& {1 \over 2 \pi} \int_0^{\infty} dk_1
[k_+ + k_- + {2 \over \beta} (log ~(1 + e^{- \beta k_+}) + log ~(1
+ e^{- \beta k_-})) - 2 k_1 \nonumber \\ &-& (m^2 + M^2) {1 \over
\sqrt{k_1^2 +
\Delta^2}}] ~.
\end{eqnarray}

\bigskip
When $m^2 = 0$, this expression gives us the effective potential at
finite temperature for the pure Cooper-pairing model considered in
ref. 1. Explicitly we have

\begin{eqnarray}
V_{eff} = {M^2 \over 4 \pi} [ln {M^2 \over \Delta^2} - 1] - {2
\over \beta} \int_0^{\infty} {dk \over \pi} log~ [1 + e^{- \beta \sqrt{k^2
+ M^2}}] ~.
\end{eqnarray}

\bigskip\noindent
Note that it is independent of the chemical potential, as was the
case at $T = 0$. For $T >> M$ we can expand the integral in eq.
(5.8) to obtain [12]

\begin{eqnarray}
V_{eff} = {M^2 \over 2 \pi} [log~ ({\pi T \over \Delta}) - \gamma]
~,
\end{eqnarray}

\bigskip\noindent
where $\gamma$ is Euler's constant. The minimum of this function
occurs at $M = 0$, which means that the condensate vanishes for
large $T$, as expected. This feature is also borne out by numerical
evaluation of eq. (5.8) [13].

\bigskip
In this paper, we have derived the general forms for the effective
potential in leading order in ${1 \over N}$, eqs. (5.4)-(5.6). We
have analyzed the case $\mu = T = 0$ in detail, showing how the
phase structure is governed by the relative magnitude of the two
constants $\alpha_1$ and $\alpha_2$. For $\mu = T = 0$ this
structure is remarkably symmetric in the two condensates $m$ and
$M$. We expect this symmetry to disappear in the general case,
however, because $\mu$ acts to suppress $m$. To study this will
require careful numerical analysis of the integrals in eqs.
(5.5)and(5.7), in order to see how the extrema of $V_{eff}$ depend
on the parameters $\alpha_1, \alpha_2, \mu$ and $T$. This work is
currently being actively pursued. It is our hope that the results
may be useful for a variety of condensed matter systems, for QCD,
and perhaps may have cosmological implications as well.


\bigskip\noindent
{\bf Acknowledgements}

We wish to thank Gregg Gallatin for interesting conversations and
Wenjin Mao for help with the numerical work. The research of AC is
supported in part by DOE grant DE-FG02-92ER-40704. In addition, AC
and HM are supported in part by the Grant-in-Aid for International
Scientific Research No. 09045036, Inter-University Cooperative
Research, Ministry of Education, Science, Sports and Culture of
Japan. This work has been performed as an activity supported by the
TMU-Yale Agreement on Exchange of Scholars and Collaborations. FC
and HM are grateful for the hospitality of the Center for
Theoretical Physics at Yale.

\bigskip\noindent
{\bf REFERENCES}

\bigskip\noindent
[1] A. Chodos, H. Minakata and F. Cooper, Phys. Lett. {\bf B449},
260 (1999).

\noindent
[2] D. Bailin and A. Love, Phys. Rep. {\bf 107} (1984) 325;
M. Iwasaki and T. Iwado, Phys. Lett. {\bf 350B} (1995) 163;
M. Alford, K. Rajagopal and F. Wilczek, Phys. Lett. {\bf 422B} (1998)
247; R. Rapp, T. Sch\"{a}fer, E.V. Shuryak and M. Velkovsky, Phys.
Rev. Lett. {\bf 81} (1998) 53.

\noindent
[3] M. Alford, K. Rajagopal and F. Wilczek, Nucl. Phys. {\bf A638},
515c (1998) and Nucl. Phys. {\bf B537}, 443 (1999); J. Berges and
K. Rajagopal, Nucl. Phys. {\bf B538}, 215 (1999); T. Sch\"{a}fer,
Nucl. Phys. {\bf A642}, 45 (1998); T. Sch\"{a}fer and F. Wilczek,
Phys. Rev. Lett. {\bf 82}, 3956 (1999).

\noindent
[4] N. Evans, S.D.H. Hsu and M. Schwetz, hep-ph/9808444 and Phys.
Lett. {\bf B449}, 281 (1999); T. Sch\"{a}fer and F. Wilczek, Phys.
Lett. B450, 325 (1999).

\noindent
[5] D.J. Gross and A. Neveu, Phys. Rev. {\bf D10} (1974) 3235.

\noindent
[6] S. Coleman, Comm. Math. Phys. {\bf 31} (1973) 259; N.D. Mermin
and H. Wagner, Phys. Rev. Lett. {\bf 17} (1966) 1133.

\noindent
[7] E. Witten, Nucl. Phys. {\bf B145} (1978) 110.

\noindent
[8] J. Hubbard, Phys. Rev. Lett. {\bf 3}, 77 (1959); R.L.
Stratonovich, Doklady Akad. Nauk. SSSR {\bf 115}, 1097 (1957); S.
Coleman, {\it Aspects of Symmetry}, Cambridge Press, 1985, p. 354.

\noindent
[9] U.  Wolff, Phys. Lett. {\bf B 157}, 303 (1985).

\noindent
[10] T. Inagaki, in Proceedings of the 4th Workshop on Thermal
Field Theories and their Applications, Dalian, China, 1995, pp.
121-130 (hep-ph 9511201).

\noindent
[11] A. Chodos and H. Minakata, Phys. Lett. {\bf A 191}, 39 (1994),
and Nucl. Phys. {\bf B 490}, 687 (1997).

\noindent
[12] One first computes dV/dM, then expands the integral
asymptotically in M/T, and then integrates on M. The relevant
expansion is given in L. Dolan and R. Jackiw, Phys. Rev. {\bf D9},
3320 (1974), eqs. (C13) and (C16).

\noindent
[13] Wenjin Mao, private communication.

\end{document}